# Direct observation of free excitons in luminescence spectra of xenon clusters

O.G. Danylchenko, Yu.S. Doronin, S.I. Kovalenko,

M.Yu. Libin, V.N. Samovarov[*], V.L. Vakula

*B. Verkin Institute for Low Temperature Physics and Engineering
of the National Academy of Sciences of Ukraine,
47 Lenin Ave., Kharkiv 61103, Ukraine*

**Abstract**

Luminescence of surface and free bulk excitons is detected in xenon for the first time for substrate-free rare-gas clusters. Xenon clusters were produced by the method of gas condensation in a supersonic jet emitted into vacuum. Optical study was accompanied by electron diffraction measurements to determine the structure of clusters.

**1.** Free and self-trapped excitons are fundamental electronic excitations in bulk rare-gas cryocrystals and make the main subject of the optical study of these materials. The spectrum of free excitons has been registered in absorption and luminescence measurements on bulk cryocrystals [1]. In substrate-free rare-gas clusters, the existence of excitonic levels has been confirmed by measuring absorption spectra which are formed within the characteristic time of $10^{-14}$-$10^{-15}$ s [2]. The lifetime of a free-exciton is significantly higher ($\approx 10^{-9}$ s), therefore it can efficiently dump its energy while interacting with admixtures, defects, crystal surface, as well as self-trapping into atomic and molecular excited centres. Observation of free-exciton luminescence in bulk xenon crystals required perfect defect-free samples.

---

[*] e-mail: samovarov@ilt.kharkov.ua

Until the present paper, there have been no reports of observation of luminescence from bulk and surface excitons in substrate-free rare-gas clusters, including xenon. At the same time, detection of excitonic luminescence is of interest not only to the physics of cryocrystals, but opens wide prospective on the study of new quantum effects, including quantum dots based on van der Waals clusters. Recently, quantum-confinement effects have been widely studied mostly on semiconductor clusters and nanocrystallites, e.g., by luminescence spectra of free excitons in CdS and similar systems [3].

**2.** We used the method of gas condensation in a supersonic jet flowing into vacuum [4]. Mixtures of argon and xenon were studied, the xenon concentration $C$ in the original gas mixtures was varied from 1 through 6.5% (for concentrations less than 1% see Ref. 4). The pressure $p_0$ and temperature $T_0$ at the nozzle inlet were controlled in the range from 0.5 through 2.5 atm and from 240 through 160 K, respectively. Electron diffraction experiments combined with optical measurements have shown that for various values of concentration, temperature, and pressure the supersonic jet can consist mainly of: (i) mixed argon-xenon clusters having a xenon icosahedral (quasicrystalline) core; (ii) pure xenon icosahedral clusters; (iii) mixed argon-xenon clusters with a crystalline xenon core; (iv) pure crystalline xenon clusters; (v) mixed clusters having a crystalline xenon core and an amorphous argon shell. In mixed argon-xenon clusters there is a sharp interface between the xenon core and the argon shell. In a beam composed of pure xenon clusters (cases (ii) and (iv)), the argon serves as carrier gas. The cluster temperature was 35-40 K for all the cases mentioned above. The radiation spectra in the region of 8.1-8.5 eV were excited by a 1 keV electronic beam.

**3.** Depending on the initial set of parameters, we could observe rich spectra from bulk and surface excitons. Here we report one of them corresponding to the case with $C = 3\%$, $p_0 = 1$ atm, $T_0 = 165$ K. Fig. 1 shows the electron diffraction pattern (*a*) and the luminescence spectrum (*b*). The diffraction pattern displays only the peaks which are related to the crystalline xenon (in particular, there are no peaks (311) at the diffraction vector value s=3.75 Å$^{-1}$ and (200) at 2.37 Å$^{-1}$ which are characteristic of argon). According to the estimations based on diffraction maxima halfwidths, the size of xenon clusters was no less than 1500 atoms (the cluster diameter being about 55 Å, which corresponds to nearly 7 occupied Mackay spheres).

The excitonic bands are red-shifted with respect to the atomic line $^3P_1$ (8.44 eV), which corresponds to the radiation of atoms desorbed from clusters. It should be noted that the desorption of these excited atoms was not observed in bulk xenon [5]. The 8.35 eV band was assigned to the radiation of free excitons from the cluster volume (in bulk cryocrystals this band was observed repeatedly at such energies). At lower frequencies there is a wide non-elementary feature composed of three bands at 8.31, 8.29, and 8.26 eV. We could observe one, two, or three of the components with various intensity ratio between them depending on the initial set of the experimental parameters. The 8.31 and 8.25 eV peaks were observed earlier in bulk xenon [6] and assigned to the Xe n=1 surface exciton components due to the crystal field splitting.

Fig. 2 demonstrates another situation which takes place for $p_0$=2 atm, $T_0$=170 K, $C$=6.5%. As can be seen from the electron diffraction pattern (Fig. 2$a$), the xenon diffraction maxima are accompanied by the argon peak (311) at s=3.75 Å$^{-1}$. The maximum at s=2.00 Å$^{-1}$ is very prominent due to the overlapping argon and xenon diffraction peaks. This pattern corresponds to crystalline xenon clusters covered with a thin argon shell. Fig. 2$b$ shows that the luminescence spectrum lacks the bulk free exciton band and the central peak at 8.29 eV of the surface exciton maximum. The other surface exciton peaks at 8.25 and 8.31 eV are rather intensive. It can be concluded that the thin argon shell suppresses the bulk excitons.

In qualitative approximation, the observed spectra can be explained as follows. As is known, the bulk exciton with n=1 has a radius of 3.2 Å in bulk xenon, which is smaller than the Xe nearest-neighbour distance (4.3 Å). The mean free path of the exciton is several hundreds Ångströms and exceeds noticeably the cluster size. There are following channels of inelastic scattering for a free exciton: (i) formation of a two-centre state $Xe_2^*$ with a self-trapping barrier (we observed the radiation from both bulk excitons and $Xe_2^*$ at ≈7.1 eV); (ii) transformation of a bulk exciton into surface modes; (iii) energy dumping accompanied by desorption of excited and neutral atoms. The observation of bulk excitons provides evidence for an elastic scattering channel at the interface with vacuum during the excitonic lifetimes of $10^{-9}$ s with rather high probability of exciton reflection from the interface. For the case of the xenon-argon interface, an inelastic scattering of bulk excitons takes place resulting in an almost complete conversion of energy into luminescence centres of another nature.

The complete set of our experimental results as well as their analysis will be published in *Low Temperature Physics* (2007) and elsewhere.

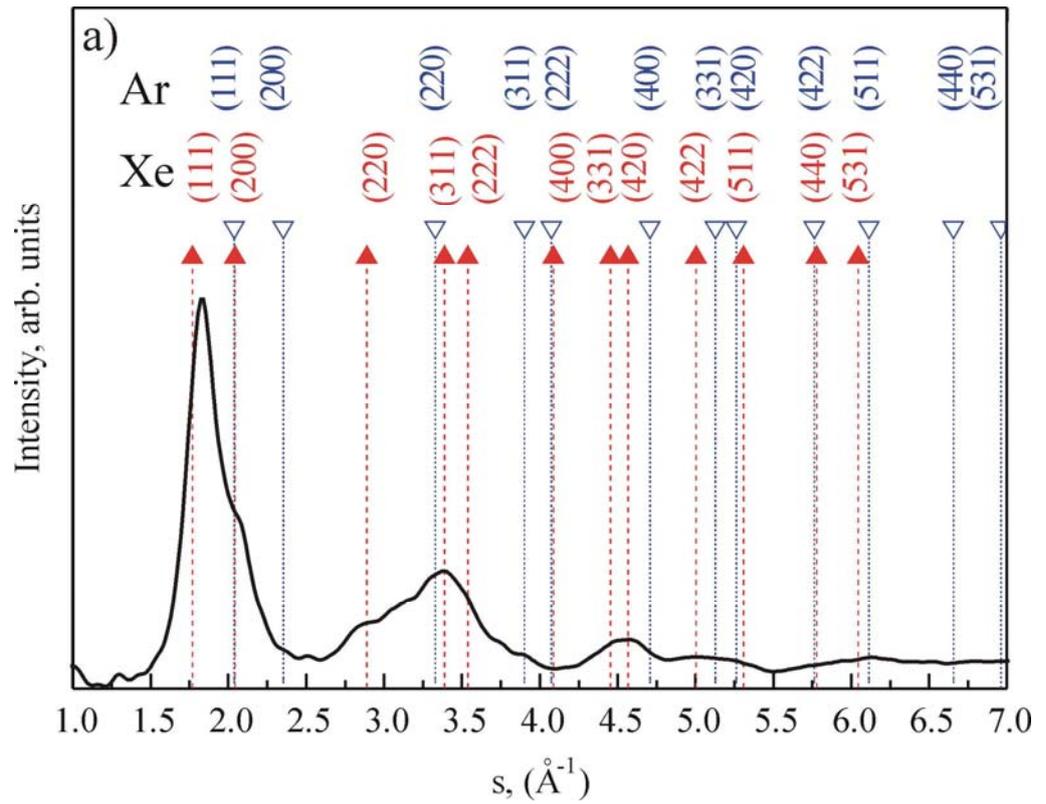

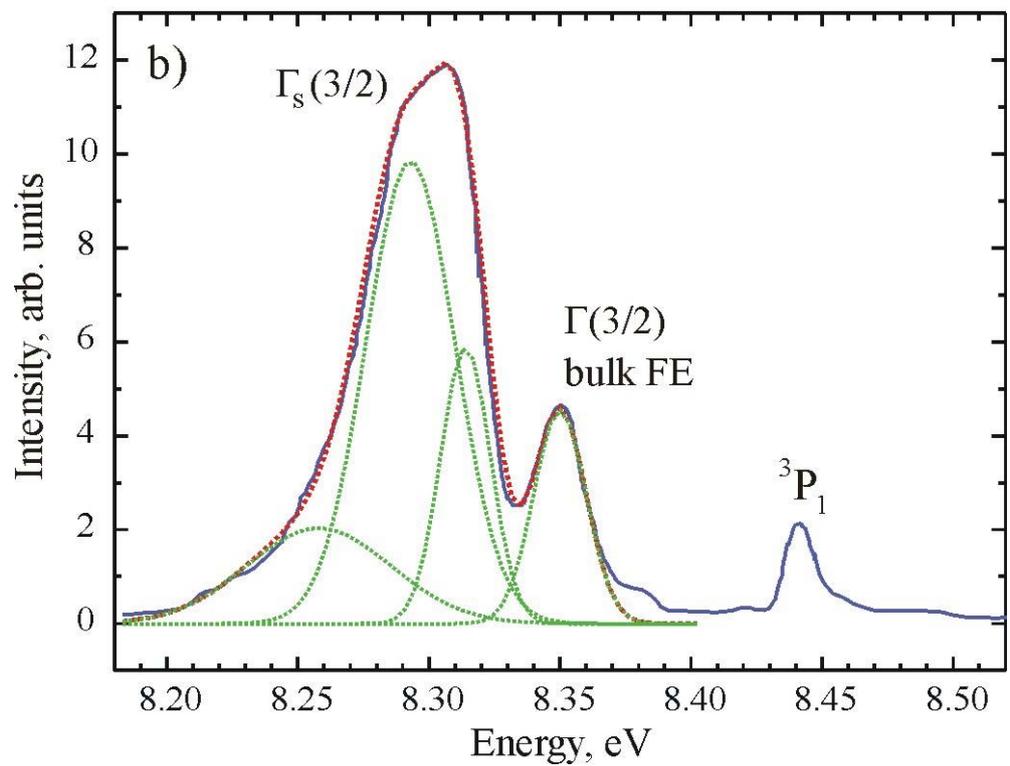

Fig. 1. Electron diffraction pattern (*a*) and cathodoluminescence spectrum (*b*) of substrate-free xenon clusters in gaseous argon.

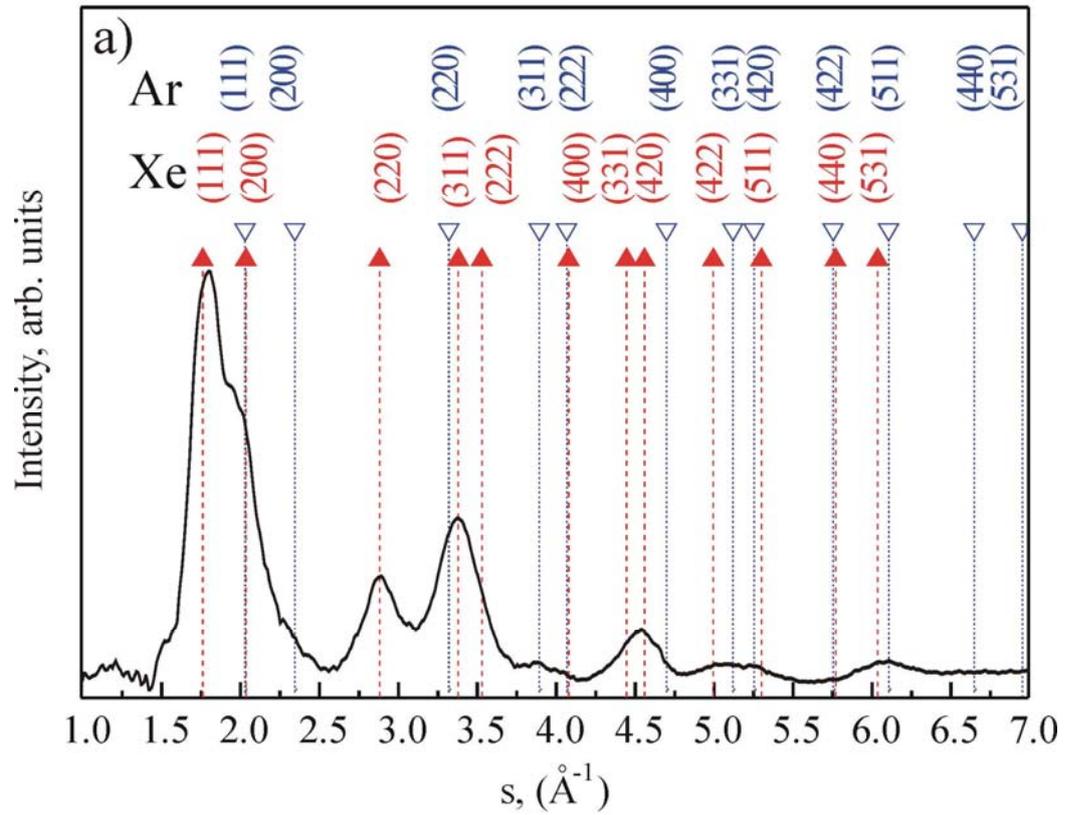

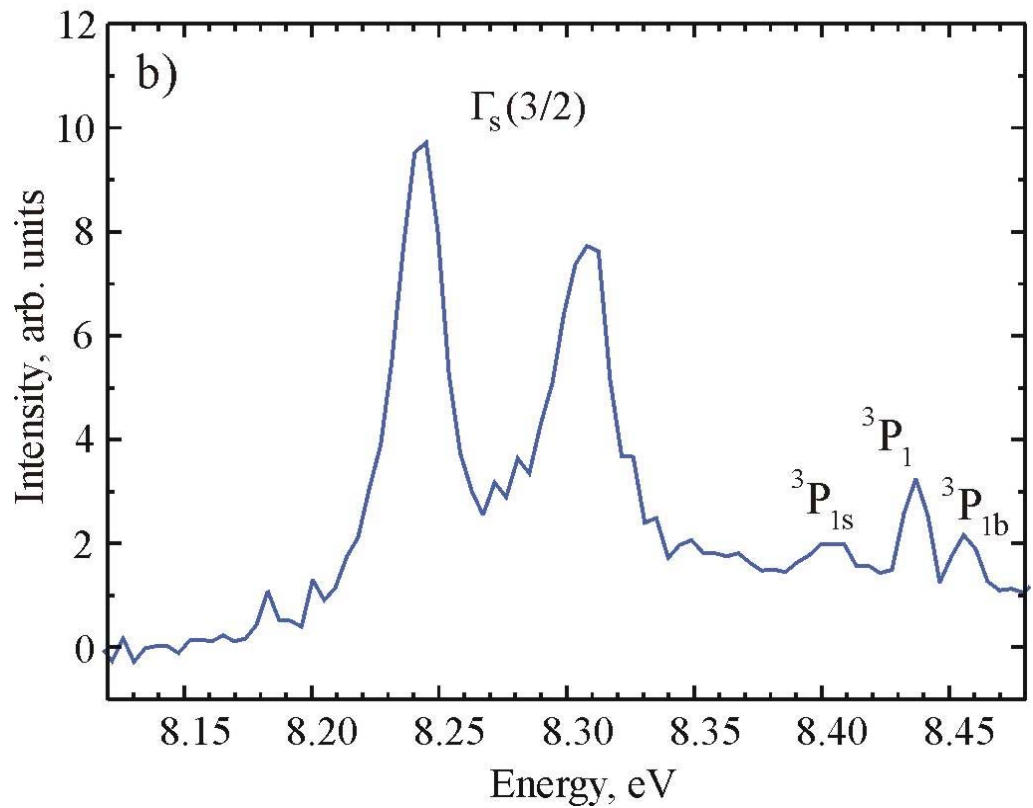

Fig. 2. Electron diffraction pattern (*a*) and cathodoluminescence spectrum (*b*) of substrate-free xenon clusters covered with a thin argon shell.